\documentclass{elsart5p}

\usepackage{amssymb,amsmath} 
\usepackage{bm,hyphenat,xspace} 
\usepackage{graphicx,epsfig}

\newcommand{\A}[2]{{}^{#1}\mathrm{#2}}

\newcommand{\clh}{\mathcal{H}}

\newcommand {\mbf}[1]{{\mathbf{#1}}}

\newcommand {\vecg}[1]{\mbox{\boldmath{$#1$}} }
\newcommand{\cm}{\mathrm{c\!\:\!.m\!\:\!.}}
\newcommand{\email}[1]{\ead{#1}}

\begin{document}


\begin{frontmatter}

\title {Core-excitation effects in ${}^{20}\mathrm{O}(d,p){}^{21}\mathrm{O}$ 
transfer reactions: \\
Suppression or enhancement?}

\author{A.~Deltuva}, 
\email{arnoldas.deltuva@tfai.vu.lt}
\author{D. Jur\v{c}iukonis},
\email{darius.jurciukonis@tfai.vu.lt}
\author{E. Norvai\v{s}as}
\email{egidijus.norvaisas@tfai.vu.lt}
\address{
Institute of Theoretical Physics and Astronomy, 
Vilnius University, Saul\.etekio al. 3, LT-10222 Vilnius, Lithuania}



\begin{abstract}
${}^{20}\mathrm{O}(d,p){}^{21}\mathrm{O}$ transfer reactions are described 
using  momentum-space Faddeev-type equations for transition operators and
including the vibrational excitation of the ${}^{20}\mathrm{O}$ core. 
The available experimental  cross section data at 10.5 MeV/nucleon beam energy for the 
  ${}^{21}\mathrm{O}$  ground state $\frac52^+$ and excited state $\frac12^+$ 
are quite well reproduced by our calculations including the core excitation.
Its effect can be roughly simulated
reducing the single-particle cross section by the corresponding spectroscopic factor.
Consequently, the extraction of the spectroscopic factors taking the
ratio of experimental data and single-particle cross section at this energy
 is a reasonable procedure. However, at higher energies
core-excitation effects are much more complicated and have no simple
relation to spectroscopic factors. We found that core-excitation effects
are qualitatively very different for reactions with the orbital angular momentum
transfer $\ell=0$ and $\ell=2$,
suppressing the cross sections for the former and enhancing for the latter,
and changes the shape of the angular distribution in both cases.
Furthermore, the core-excitation effect is a result of a complicated interplay between its 
contributions of the two- and three-body nature.
\end{abstract}


\end{frontmatter}

\section{Introduction \label{sec:intro}}

Interactions between nucleons ($N$) and composite nuclei ($A$) are usually modeled
by two-body effective optical or binding potentials acting between structureless particles.
 This scheme works quite well for stable tightly bound nuclei 
but may become a poor approximation for exotic nuclei that nowadays are extensively studied
both experimentally and theoretically. An improvement of the structureless nucleus model,
at a first step, consists in explicitly considering also its lowest excited states ($A^*$),
thereby accounting for the compositeness of the nucleus $A$ in an approximate way.
This extension has been proposed long ago \cite{tamura:cex} and applied to numerous studies
of elastic and inelastic $N+A$ scattering. However, the application of interaction models 
including the excitation of the involved nucleus, also called the core excitation,
 to three-body nuclear reactions,
e.g., deuteron ($d$) stripping and pickup, is still a complicated task. 
First studies of $(d,p)$ reactions demonstrating the importance of the core excitation
\cite{PhysRev.181.1396,Glendenning1971575,Mackintosh1971353,Ascuitto1974454} 
were based on two-body-like
approaches such as the distorted-wave Born approximation (DWBA)
and coupled-channels Born approximation (CCBA)
that relied on deuteron-nucleus optical potentials.
Only quite recently the three-body calculations have emerged that include the core excitation.
Extensions of the DWBA
 \cite{moro:12a,moro:12b} 
and continuum discretized coupled channels (CDCC)  method \cite{summers:07a,diego:14a}
mostly focused on the breakup reactions, in particular,
 of  $\A{11}{Be}$. The calculations for neutron transfer
reactions  $\A{10}{Be}(d,p)\A{11}{Be}$ and $\A{11}{Be}(p,d)\A{10}{Be}$ were performed using 
 rigorous Faddeev three-body  scattering theory \cite{faddeev:60a} 
in the form of Alt, Grassberger, and Sandhas (AGS) equations \cite{alt:67a} for transition operators, 
solved in the extended Hilbert space \cite{deltuva:13d,deltuva:15b,deltuva:16c}.
The latter works demonstrated that in the deuteron stripping and pickup the core excitation effect 
cannot be simply simulated by the reduction of the cross section 
according to the respective spectroscopic factor (SF). It was found that extracting the SF from the ratio
of experimental and theoretical  transfer cross sections, as often used with the
adiabatic distorted wave approximation (ADWA) calculations \cite{johnson:70a}, 
may lead to a strong underestimation of the SF.
Calculations of Refs.~\cite{deltuva:13d,deltuva:15b,deltuva:16c} employed the rotational model \cite{tamura:cex}
for the excitation of the $\A{10}{Be}$ core; the most prominent core-excitation effects
have been observed  for the $\A{10}{Be}(d,p)\A{11}{Be}$ transfer to the ground state of
$\A{11}{Be}(\frac12^+)$ whose dominant component corresponds to an $S$-wave neutron coupled
to the $\A{10}{Be}(0^+)$ ground state, i.e., the orbital angular momentum transfer 
for this reaction is $\ell=0$.
 In contrast, for the $\ell=1$ transfer leading to the excited state
$\A{11}{Be}(\frac12^-)$ the core-excitation effects have been less remarkable.
It is therefore very important to clarify the systematics of the core-excitation effects
in transfer reactions, investigating
 other types of excitation mechanisms and bound states. Furthermore, a deeper understanding may be
gained by disentangling the effects of two- and three-body nature.
The study of $\A{20}{O}(d,p)\A{21}{O}$ transfer reactions intended in the present work
leads to the desired goal and is interesting for several reasons. 
First, the $\A{21}{O}(\frac52^+)$ ground state 
has a significant component of $D$-wave neutron coupled
to the $\A{20}{O}(0^+)$ ground state, thereby allowing the extension of systematics
from Refs.~\cite{deltuva:13d,deltuva:15b,deltuva:16c} to  the $D$-wave neutron state and  $\ell=2$ transfer.
Second, the lowest excitation of the $\A{20}{O}$ core $2^+$ has a vibrational character,
giving opportunity to investigate the vibrational model for the nucleon-core interaction
\cite{tamura:cex} in the context of  transfer reactions.
Last but not least there are experimental data for  $\A{20}{O}(d,p)\A{21}{O}$ transfer reactions
 at 10.5 MeV/nucleon beam energy \cite{o20dp}
that have not yet been analyzed with rigorous Faddeev-type calculations.

In Sec.~\ref{sec:2} we shortly recall the three-body scattering equations
with core excitation, and in Sec.~\ref{sec:3}
describe the employed  nucleon-$\A{20}{O}$ potentials.
Results are presented in  Sec.~\ref{sec:4},
and a summary is given in  Sec.~\ref{sec:5}.

\section{Solution of three-body scattering equations with core excitation 
\label{sec:2}}

The numerical technique for calculating deuteron-nucleus reactions with the inclusion
of the core excitation is taken over from Refs.~\cite{deltuva:13d,deltuva:15b,deltuva:16c} but
further developments are needed to get insight into separate core-excitation contributions
of the two- and three-body nature.
The method is based on the integral formulation of rigorous Faddeev-type three-body scattering theory
for transition operators as proposed by 
 Alt, Grassberger, and Sandhas \cite{alt:67a}, but extended for the Hilbert 
space $\clh_g \oplus \clh_x$ whose sectors correspond to the core being in
its ground (g) or excited (x) state.  
These sectors are coupled by the
nucleon-core two-body potentials $v_\alpha^{ji}$ where the 
superscripts $j$ and $i$, being either $g$ or $x$, label the internal states of the core,
and the subscript $\alpha$, being $A$, $p$, or $n$, labels the spectator particle
in the odd-man-out notation.
Consequently, the respective two-body transition operators
\begin{equation}  \label{eq:Tg}
T_{\alpha}^{ki} =  v_{\alpha}^{ki} +\sum_{j=g,x} 
v_{\alpha}^{kj} G_0^{j} T_{\alpha}^{ji}
\end{equation}
and three-body transition operators
\begin{equation}  \label{eq:Uba}
U_{\beta \alpha}^{ki}  = \bar{\delta}_{\beta\alpha} \, \delta_{ki} {G^{i}_{0}}^{-1}  +
\sum_{\gamma=A,p,n} \sum_{j=g,x}   \bar{\delta}_{\beta \gamma} \, T_{\gamma}^{kj}  \,
G_{0}^{j} U_{\gamma \alpha}^{ji}
\end{equation}
couple $\clh_g$ and  $\clh_x$ as well. 
Here $\bar{\delta}_{\beta\alpha} = 1 - \delta_{\beta\alpha}$ and
 $G_0^{j} = (E+i0-\delta_{jx}\Delta m_A - K)^{-1}$ is the projection of
the free resolvent into  $\clh_j$, with $E$, $\Delta m_A$,  and $K$ being
the available energy in the center-of-mass (c.m.) frame,  
core-excitation energy, and kinetic energy operator, respectively.
The amplitudes for deuteron stripping reactions $A(d,p)B$, $B$ denoting
the $(An)$ bound state, are given by the on-shell matrix elements
$\langle \Phi_p^g|U_{pA}^{gg}|\Phi_A^g \rangle + \langle \Phi_p^x|U_{pA}^{xg}|\Phi_A^g \rangle$ 
since the final $p+B$ channel state  
$|\Phi_p \rangle = |\Phi_p^g \rangle + |\Phi_p^x \rangle$ has components in both
Hilbert sectors. 

\begin{figure}[!]
\begin{center}
\includegraphics[scale=0.6]{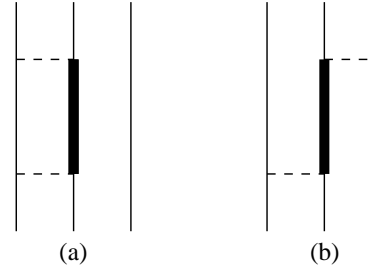}
\end{center}
\caption{\label{fig:cex23}
Diagrammatic representation of the lowest-order core-excitation
contributions of (a) two-body and (b) three-body nature.
Horizontal dashed lines stand for potentials while vertical solid lines stand for particles,
the thick one being for the core in its excited state.}
\end{figure}

The core-excitation effects can be separated into contributions of two- and
and three-body nature. The former consists in modifying 
$T_{\alpha}^{gg}$ through intermediate core excitations, i.e.,
through the terms of type $v_{\alpha}^{gx} G_0^{x} v_{\alpha}^{xg}$ and so on
 in the iterated coupled-channel Lippmann-Schwinger equation \eqref{eq:Tg}.
The contribution of the three-body nature arises due to  nondiagonal
components $T_{\alpha}^{xg}$ and $T_{\alpha}^{gx}$ that are responsible for the
coupling  of the two Hilbert sectors in Eq.~\eqref{eq:Uba}, i.e.,
$T_{\beta}^{gx} \bar{\delta}_{\beta\alpha} G_0^{x} T_{\alpha}^{xg}$ and so on, yielding,
in fact, an  energy-dependent effective three-body force (E3BF). 
Lowest-order diagrams for both types are depicted in Fig.~\ref{fig:cex23}.
We note a formal similarity between these contributions and 
the so-called dispersive and three-nucleon force effects arising
in the description of the 
three-nucleon system with the $\Delta$-isobar excitation \cite{hajduk:83a,deltuva:03c}.
Since the full core-excitation effect will be extracted from the 
solution of Eq.~\eqref{eq:Uba}, to get insight into the importance 
of separate  two- and and three-body contributions it is enough to 
exclude one of them. It is most convenient to do so for the E3BF,
whose exclusion can be achieved  by setting 
 $T_{\gamma}^{kj} = \delta_{kg}  \delta_{jg} T_{\gamma}^{gg}$ in  Eq.~\eqref{eq:Uba}.
This type of results will be labeled in the following as ``no E3BF''.

Although the present work employs the potentials $v_\alpha^{ji}$ derived from
the vibrational model \cite{tamura:cex}, calculations proceed in the same way as 
with rotational model potentials used in  Refs.~\cite{deltuva:13d,deltuva:15b,deltuva:16c}. 
The AGS equations \eqref{eq:Uba} are solved numerically in the momentum-space 
partial-wave representation. Six sets of base functions
$|p_\alpha q_\alpha 
(l_\alpha \{ [ L_\alpha (s^i_\beta s^i_\gamma)S^i_\alpha] j^i_\alpha s^i_\alpha\} \mathcal{S}^i_\alpha)
J M \rangle $ are employed with
 $(\alpha\beta\gamma) = (Apn)$, $(pnA)$, or $(nAp)$, and $i = g$ or $x$. 
Here $p_\alpha$ and  $q_\alpha$ are magnitudes of Jacobi momenta for the configuration
$\alpha(\beta\gamma)$ while  $L_\alpha$ and  $l_\alpha$ are the associated orbital
angular momenta. Furthermore,
 $s^i_A$ and $s^i_p = s^i_n = \frac12$ are spins of the corresponding particles,
among them only $s_A^i$ depends on the Hilbert sector $i$, i.e.,
 $s_A^g = 0$ and $s_A^x = 2$ 
in the considered case of the $\A{20}{O}$ nucleus with the ground and first excited
states $0^+$ and $2^+$, respectively.
All discrete angular momentum quantum numbers, via the 
intermediate angular momenta $S^i_\alpha$, $j^i_\alpha$, and $ \mathcal{S}^i_\alpha$,
 are coupled to the total angular momentum $J$ with the projection $M$.
We note that the spin $s_A^x = 2$ implies roughly five times more basis
states in $\clh_x$  as compared to 
$\clh_g$, thereby increasing the demand on computer memory 
and time by a factor of 20 to 40. Including more states of the core, e.g., 
the second excited state $4^+$ would be even significantly more demanding, and for this reason
we restrict our present calculations to the inclusion of $0^+$ and $2^+$ states
of $\A{20}{O}$. 
 Well-converged results for  $\A{20}{O}(d,p)\A{21}{O}$ transfer reactions are obtained by
including $J \le 25$ states
with $L_A \le 3$, $L_p \le 5$, and  $L_n \le 10$. Higher value for $L_n$ is needed due to
the Coulomb force present within the $A+p$ pair which is 
included via the screening and renormalization
method \cite{taylor:74a,alt:80a,deltuva:05a}. 

\section{Potentials \label{sec:3}}

\begin{figure}[!]
\begin{center}
\includegraphics[scale=0.69]{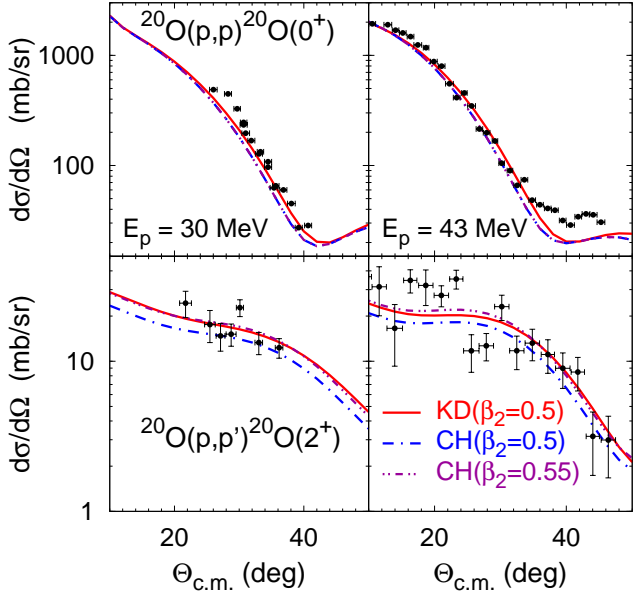}
\end{center}
\caption{\label{fig:o20p}  (Color online)
Differential cross sections  $d\sigma/d\Omega$ for  elastic (top) and
inelastic (bottom) $p+\A{20}{O}$ scattering at 30 (left) and 43 (right) MeV/nucleon
 beam energies as functions of the c.m. scattering angle $\Theta_\cm$.
Results  including the core excitation based
on KD and CH potential models with  $\beta_2 = 0.5$ and $0.55$
are compared with the experimental data from Ref.~\cite{o20p30}
(30 MeV) and Ref.~\cite{o20p43} (43 MeV).}
\end{figure}

We consider the system of a proton, a neutron, and a $\A{20}{O}$ core
with masses $m_p = 0.99931\,m_N$, $m_n = 1.00069\,m_N$, and
 $m_A = 19.84153\,m_N$ given in units of 
  $m_N = (m_n + m_p)/2 = 938.919$ MeV; the core excitation energy is $\Delta m_A = 1.684$ MeV.
To the best of our knowledge,
potentials specifically designed for the $N$+$\A{20}{O}$ interaction including the
core excitation are not available. The corresponding experimental data are scarce,
we are  aware of only two  $p$+$\A{20}{O}$ elastic and inelastic scattering
measurements at 30 \cite{o20p30} and 43 \cite{o20p43} MeV/nucleon beam energies. 
In these works the data have been analyzed using DWBA or coupled-channel
calculations with global optical potentials, e.g., \cite{becchetti}.
Extracted values of the quadrupole vibrational coupling parameter $\beta_2$ are
$0.50 \pm 0.04$  \cite{o20p30} and $0.55 \pm 0.06$ \cite{o20p43}.
We also  base our calculations on  global optical potentials but use more modern
parametrizations, namely, those
of Koning-Delaroche (KD) \cite{koning} and Chapel Hill 89 (CH) \cite{CH89}.
These potentials were designed for $A \ge 24$ and $A \ge 40$ nuclei, respectively,
but one may expect a reasonable extrapolation also to $A = 20$, especially for
the KD potential.
To include the core excitation, we extended these potentials 
for quadrupole vibrations \cite{tamura:cex} and
modify by the subtraction method of Ref.~\cite{deltuva:15b}
 adding a nonlocal contribution. The terms
up to the second order in $\beta_2$ as given in Ref.~\cite{tamura:cex}
are taken into account in our calculations.
It turns out that such an approach  reproduces the experimental
data for elastic and inelastic differential cross sections
of   Refs.~\cite{o20p30,o20p43} reasonably well using the same value $\beta_2 = 0.5$
as shown in Fig.~\ref{fig:o20p}, especially for the  KD potential.
To study the sensitivity to $\beta_2$, we also show CH predictions with
$\beta_2 = 0.55$, that yield a better  description of the inelastic cross section.
The observed agreement encourages the application of these potentials for
$\A{20}{O}(d,p)\A{21}{O}$ transfer reactions, not only for  $p$+$\A{20}{O}$ 
but also for  $n$+$\A{20}{O}$ pair where no experimental scattering data are available.
An exception is the $n$+$\A{20}{O}$ potential in the $\frac52^+$ and $\frac12^+$ partial waves 
that must be real to support bound states with the binding energies of
3.806 and 2.586 MeV, respectively. In addition,  predictions of various shell models
\cite{warburton:92a,PhysRevC.60.054315} for SF's of these 
states are available, being around 0.33 to 0.34 for $\frac52^+$
and 0.81 to 0.83 for $\frac12^+$  \cite{o20dp}. We include this information 
in constraining the  $n$+$\A{20}{O}$ potentials.
We start with the undeformed coordinate-space potential
\begin{gather}  \label{eq:Vb}
\begin{split}
v_{\alpha}(r) = & {} -V_c f(r,R,a) + \mbf{L}^2 V_L f(r,R,a)  \\
& + \vecg{\sigma}\cdot \mbf{L} \,
V_{so} \, \frac{2}{r} \frac{d}{dr}f(r,R,a),
\end{split}
\end{gather}
where $f(r,R,a) = [1+\exp((r-R)/a)]^{-1}$ is Woods-Saxon form factor,  $a=0.65$ fm,  
$V_{so}= 6.0 \, \mathrm{MeV}\cdot\mathrm{fm}^2$, and $R$ is taken from
the real part of the optical potential acting in other waves, i.e.,
$R = 3.13$ fm (3.17 fm) for KD (CH) potentials. In addition to standard central and spin-orbit 
terms a phenomenological $\mbf{L}^2$ term is taken over from Ref.~\cite{amos:03a}.
The core excitation is included by quadrupole vibrations
of the central part in \eqref{eq:Vb} with $\beta_2 = 0.5$  or 0.55 as 
described by Tamura \cite{tamura:cex}. Potential strength parameters
$V_c$ and $V_L$ are adjusted to reproduce the desired binding energies and SF's.
The latter are chosen to be the middle values of several shell model predictions \cite{o20dp},
i.e.,  0.34 for $\frac52^+$ and 0.82 for $\frac12^+$. Deeply-bound Pauli forbidden states
are projected out. The resulting potential 
parameters are collected in Tables \ref{tab:Vd} and  \ref{tab:Vs}; parameter sets with
$\beta_2 = 0.0$ correspond to single-particle models without core excitation
that are used to isolate its effect.

\begin{table} [!]
\caption{\label{tab:Vd}
Quadrupole vibration parameter $\beta_2$, Woods-Saxon radius $R$, potential strengths
$V_c$ and $V_L$, and the resulting SF for the $\A{21}{O}$ ground state $\frac52^+$
with the binding energy of 3.806 MeV.}
\begin{tabular}{l*{4}{l}}
 $\beta_2$ & $R$(fm) &  $V_c$(MeV) & $V_L/V_c$ & SF \\ \hline
 0.50 & 3.13 & 53.564 & 0.0389 & 0.34 \\
 0.50 & 3.17 & 52.580 & 0.0396 & 0.34 \\
 0.55 & 3.17 & 51.907 & 0.0419 & 0.34 \\  \hline
 0.0  & 3.19 & 50.425 & 0.0    & 1.0 \\
 0.0  & 3.23 & 49.347 & 0.0    & 1.0
\end{tabular}
\end{table}

\begin{table} [!]
\caption{\label{tab:Vs}
Quadrupole vibration parameter $\beta_2$, Woods-Saxon radius $R$, potential strengths
$V_c$ and $V_L$, and the resulting SF for the $\A{21}{O}$ excited state $\frac12^+$
with the binding energy of 2.586 MeV.}
\begin{tabular}{l*{4}{l}}
 $\beta_2$ & $R$(fm) &  $V_c$(MeV) & $V_L/V_c$ & SF \\ \hline
 0.50 & 3.13 & 45.531 & 0.0252 & 0.82 \\
 0.50 & 3.17 & 44.639 & 0.0260 & 0.82 \\
 0.55 & 3.17 & 44.038 & 0.0308 & 0.82 \\  \hline
 0.0  & 3.19 & 49.743 & 0.0    & 1.0 \\
 0.0  & 3.23 & 48.813 & 0.0    & 1.0
\end{tabular}
\end{table}

\section{Results \label{sec:4}}

\begin{figure}[!]
\begin{center}
\includegraphics[scale=0.69]{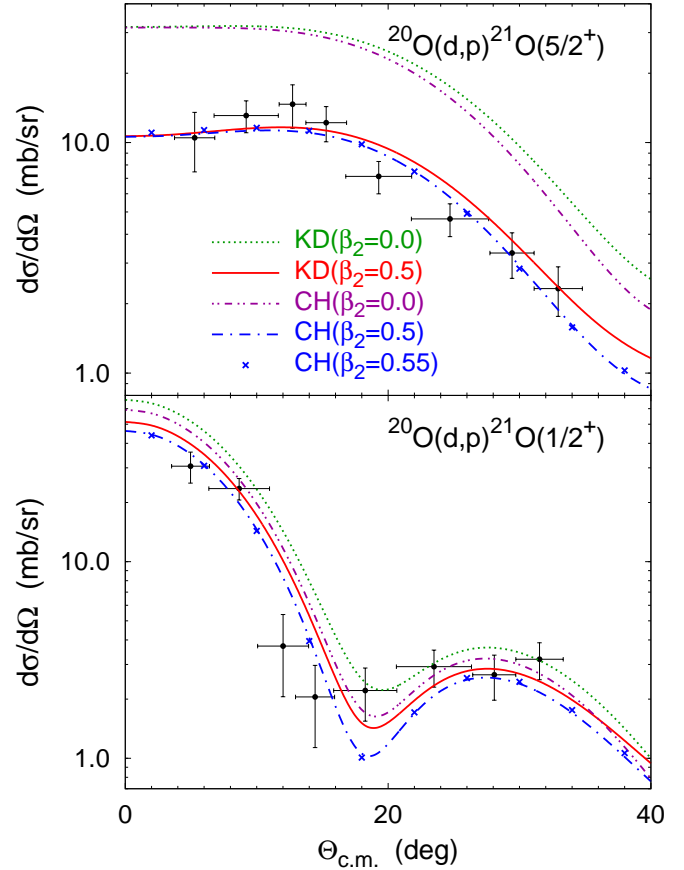}
\end{center}
\caption{\label{fig:dop}  (Color online)
Differential cross section  for  $\A{20}{O}(d,p)\A{21}{O}$ transfer reactions
at  $E_d = 21$ MeV leading to $\A{21}{O}$ ground $\frac52^+$ (top) and
excited $\frac12^+$ (bottom) states.
 Predictions obtained with and without the vibrational core excitation based 
on KD and CH potential models 
are compared with the experimental data from Ref.~\cite{o20dp}.}
\end{figure}

Taking  $p$+$\A{20}{O}$ and  $n$+$\A{20}{O}$ potentials from previous section together
with the high-precision charge-dependent (CD) Bonn $n+p$ potential 
\cite{machleidt:01a} as the dynamic input, we solve the AGS equations  \eqref{eq:Uba}
and calculate $\A{20}{O}(d,p)\A{21}{O}$  differential cross sections $d\sigma/d\Omega$
as functions of the c.m. scattering angle $\Theta_\cm$. 
We start with  10.5 MeV/nucleon beam energy, corresponding to the deuteron beam
energy $E_d = 21$ MeV, where the 
experimental data  \cite{o20dp} are available. The results obtained without ($\beta_2 = 0$)
and with ($\beta_2 = 0.5$) core excitation based on KD and CH potentials 
are presented in  Fig.~\ref{fig:dop}.
 The core excitation effect for the transfer to the
$\A{21}{O}$ ground state $\frac52^+$ is very large. It strongly reduces 
the differential cross section bringing it in a good agreement with the
experimental data. The sensitivity to the potential model is visible
except at very small angles but remains smaller than experimental error bars.
To study the sensitivity to $\beta_2$ we include also CH-based predictions with 
$\beta_2 = 0.55$; they are almost indistinguishable from the corresponding
$\beta_2 = 0.5$ results, indicating that the value of $\beta_2$ is not critical
for transfer observables provided that other properties are fixed.
Same conclusions regarding the sensitivity to $\beta_2$ and potential 
apply also for the transfer to the $\A{21}{O}$ excited state $\frac12^+$.
However, in this case the core excitation effect is  smaller, although it also
reduces the differential cross section bringing it closer to the data,
except for few points at larger angles. There is also some mismatch between
predicted and measured positions of the minimum. We note that for both reactions
KD predictions are slightly higher, possibly due to a larger elastic 
$N$+$\A{20}{O}$ cross section.

\begin{figure}[!]
\begin{center}
\includegraphics[scale=0.6]{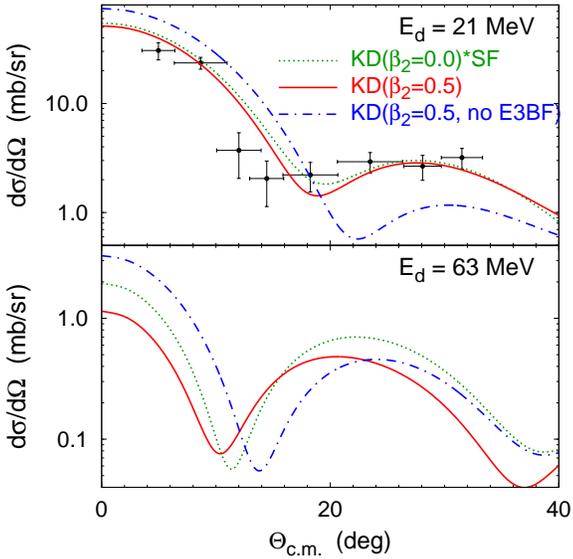}
\end{center}
\caption{\label{fig:se}  (Color online)
Differential cross section  for  $\A{20}{O}(d,p)\A{21}{O}$ transfer reactions
at  $E_d = 21$ and 63 MeV leading to $\A{21}{O}$
excited $\frac12^+$  state.
Single-particle predictions scaled by  SF $=0.82$ (dotted curves) are compared with
results including the core excitation in full (solid curves) and 
excluding the E3BF contribution (dash-dotted curves).
The experimental data at  $E_d = 21$ MeV are from Ref.~\cite{o20dp}.}
\end{figure}

\begin{figure}[t]
\begin{center}
\includegraphics[scale=0.6]{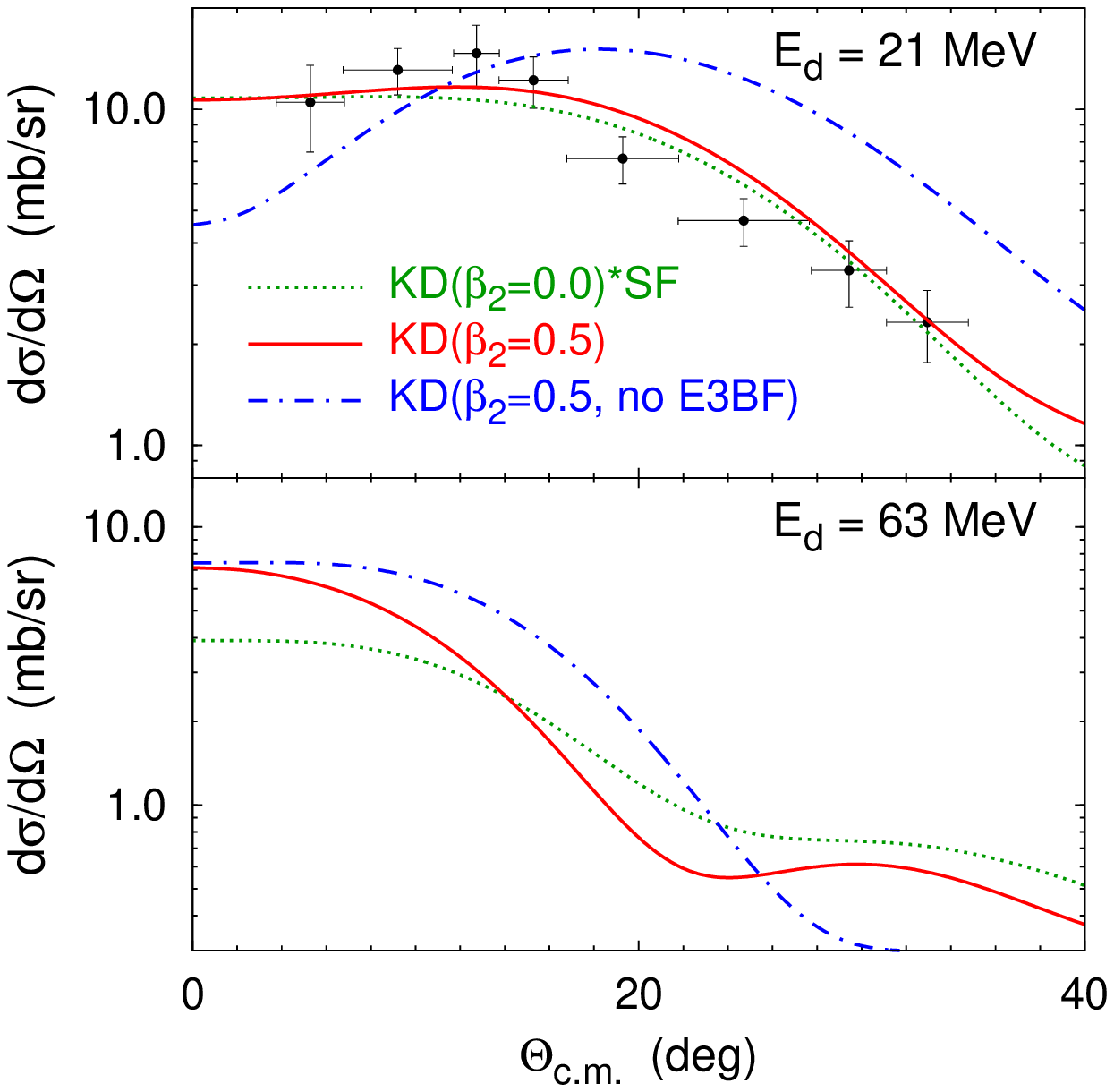}
\end{center}
\caption{\label{fig:de}  (Color online)
Same as  Fig.~\ref{fig:se} but for $\A{21}{O}$ ground state $\frac52^+$
with  SF $=0.34$. }
\end{figure}

Obviously, the reduction of the differential cross section due to the
core excitation correlates with the reduction of the SF from unity to
0.34  and 0.82 for ground and excited states, respectively.
In naive reaction methods like DWBA or ADWA the dynamic core excitation
is usually neglected, i.e., it is assumed  that the bound state component
$| \Phi_p^x \rangle$ takes no part in the reaction, and the core excitation
effect is a reduction of the single-particle differential cross section by the SF.
However,  this conjecture on factorization may be wrong as it was
demonstrated by rigorous Faddeev-type calculations  using the
 $\A{10}{Be}(d,p)\A{11}{Be}$ transfer to the ground state of
$\A{11}{Be}(\frac12^+)$ as example \cite{deltuva:13d,deltuva:15b,deltuva:16c}.
We therefore investigate in  Figs.~\ref{fig:se} and \ref{fig:de} 
the validity of factorization conjecture for 
 $\A{20}{O}(d,p)\A{21}{O}$ reactions over a broader energy range.
Having no more experimental data, we simply take additional energy value
larger by a factor of  3, i.e., $E_d = 63$ MeV.
As the core excitation effects
for KD and CH turn out to be quite similar, we show only KD results that in general
are closer to the experimental two- and three-body data. 
We multiply KD single-particle $\beta_2 = 0$ differential cross sections
by the respective SF of the model with the core excitation and compare
with the KD($\beta_2 = 0.5$) results fully including the core excitation.
The difference between these two results, or the deviation of the ratio
\begin{equation}
R_x = \frac{d\sigma/d\Omega(\beta_2=0.5)}{\mathrm{SF}\cdot d\sigma/d\Omega(\beta_2=0)}
\end{equation}
from unity indicates violation of the factorization conjecture.
We start with the excited state $\frac12^+$ analysis in 
 Fig.~\ref{fig:se} where we expect some similarities with the 
$\A{11}{Be}(\frac12^+)$ case \cite{deltuva:13d,deltuva:15b,deltuva:16c}.
At $E_d = 21$ MeV the two curves are close but, at least below the first minimum,
 differ by a roughly constant factor, i.e., the core excitation effect is slightly,
by about 6\%, stronger than predicted by the factorization conjecture.
Having the SF of 0.82 the  core excitation reduces the  differential cross section
at forward angles by a factor of 0.77 which is exactly the value of the SF extracted 
in Ref.~\cite{o20dp} relying on the factorization conjecture.
Thus, the dynamical core excitation model well explains a stronger reduction
of the cross section observed in Ref.~\cite{o20dp} as compared to the factorization conjecture.
The deviation between the two curves in Fig.~\ref{fig:se} increases with increasing energy,
and their ratio becomes angle-dependent, thereby indicating that
the factorization conjecture fails at higher energies. 
The reduction of the cross section at forward angles is significantly stronger 
than SF, e.g.,  $R_x = 0.59$ at $E_d = 63$ MeV and $\Theta_\cm =0^\circ$.
Such a behavior is indeed 
qualitatively consistent with findings of Refs.~\cite{deltuva:13d,deltuva:15b,deltuva:16c}
for $\A{11}{Be}(\frac12^+)$ within the rotational model.

A similar study of the  $\A{20}{O}(d,p)\A{21}{O}$ transfer to the 
 $\A{21}{O}$ ground state $\frac52^+$ is presented in  Fig.~\ref{fig:de}.
At $E_d = 21$ MeV the two curves are again close, especially at forward angles.
Thus, despite that SF $=0.34$ significantly deviates from unity,
the differential cross section including the core excitation
scales  well with SF, and at this energy the factorization conjecture
is valid. However, the situation changes  dramatically at higher energy
where the two curves deviate from each other in an angle-dependent way.
We emphasize that at forward angles this deviation is in opposite
direction as compared to the excited state $\frac12^+$, e.g., $R_x = 1.83$
at $E_d = 63$ MeV and $\Theta_\cm =0^\circ$. Thus, at higher energies
the factorization conjecture fails for the  $\A{21}{O}$ ground state $\frac52^+$
as well, but quantitatively the core excitation effect  is very different as compared
to the one for the excited state $\frac12^+$. 

In Figs.~\ref{fig:se} and \ref{fig:de} we also isolate the E3BF core-excitation effect,
given as the difference between the solid and dash-dotted curves.
Quite surprisingly, even at  $E_d = 21$ MeV it turns out to be significant.
Consequently, the core-excitation effect of the two-body nature must be significant
as well to cancel the E3BF to a large extent, especially at  $E_d = 21$ MeV,
 such that their sum  reproduces the full core-excitation effect.
We note that substantial cancellation of the corresponding
two- and three-body effects due to the $\Delta$-isobar excitation was often 
observed also in the nucleon-deuteron scattering \cite{deltuva:03c}.

We studied also sensitivity of the transfer cross sections to the
neutron-proton tensor force and $D$-state component in the deuteron.
Replacing the CD Bonn potential in the ${}^3S_1-{}^3D_1$ partial wave
by a central one reproducing deuteron binding and, roughly, $n$-$p$ 
 ${}^3S_1$ and ${}^3D_1$ phase shifts, leads to small but visible changes 
(smaller than KD - CH difference) in the cross sections. However, we do not consider
such a $n$-$p$ potential as realistic and therefore performed another test
calculation with the realistic Argonne V18 potential \cite{wiringa:95a} that has 
a stronger tensor force and a larger deuteron $D$-state probability as compared to  CD Bonn. 
In this case the differences were minor, so we conclude that uncertainties in a 
{\it realistic} $n$-$p$ force do not affect
the  $\A{20}{O}(d,p)\A{21}{O}$ transfer cross sections.

\begin{figure}[t]
\begin{center}
\includegraphics[scale=0.6]{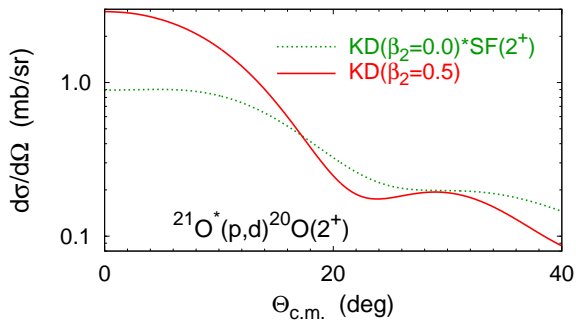}
\end{center}
\caption{\label{fig:pd}  (Color online)
Differential cross section  for  $\A{21}{O}^*(p,d)\A{20}{O}(2^+)$ transfer reactions
at  $E_p = 60.36$ MeV.
Results including the core excitation (solid curve)
are compared with single-particle predictions scaled by  SF$(2^+)=0.18$ (dotted curve).}
\end{figure}

Finally we consider the deuteron pickup reaction $\A{21}{O}(p,d)\A{20}{O}$.
For the $d+\A{20}{O}(0^+)$ final state it is exactly the time-reverse reaction
of $\A{20}{O}(d,p)\A{21}{O}$  with the cross sections
(at the same c.m. energy) related by the time reversal symmetry.
In contrast, with the $d+\A{20}{O}(2^+)$ final state it presents a new case
that we study in Fig.~\ref{fig:pd} at 60.36 MeV/nucleon beam energy.
The initial excited state $\A{21}{O}(\frac12^+)$ this time corresponds to
 the $\ell = 2$ transfer as the $\A{20}{O}(2^+)$ component is coupled with a
$D$-wave neutron. The core-excitation effect turns out to be qualitatively similar
to another $\ell = 2$ case, i.e.,  $\A{20}{O}(d,p)\A{21}{O}(\frac52^+)$
shown in the bottom part of  Fig.~\ref{fig:de}.

\section{Summary and conclusions \label{sec:5}}

We analyzed $\A{20}{O}(d,p)\A{21}{O}$ transfer reactions taking into account
the vibrational excitation of the $\A{20}{O}$ core. Calculations were performed
using Faddeev-type equations for transition operators that were
solved in the momentum-space partial-wave representation.
Well converged results were obtained
for several interaction models based on the vibrational extension
of KD and CH potentials. 

The only available experimental differential cross section data for the 
transfer to the $\A{21}{O}$ ground state $\frac52^+$ and excited state $\frac12^+$ 
at 10.5 MeV/nucleon beam energy
are quite well described by our calculations including the core excitation.
Some sensitivity to the underlying potential was observed,
but the core-excitation effects turn out to be almost independent of it.
The precise value of the quadrupole vibrational coupling  $\beta_2$ also turns
out to be irrelevant provided that spectroscopic factors are fixed
that we take from shell-model calculations.
At this lowest considered energy we found that the core-excitation effect can be approximated
to a good accuracy (6\% for the $\frac12^+$ state and even better for the $\frac52^+$ state)
 by a simple reduction of the single-particle cross section
according to the respective SF. Thus, the extraction of the SF through the
ratio of experimental data and single-particle cross section 
as performed in Ref.~\cite{o20dp} is a reasonable procedure. Nevertheless,
our prediction for a slightly stronger reduction of  the $\frac12^+$ cross section 
leads to an even better agreement between the shell model SF and experimental data.

The situation changes dramatically at higher energy where the
core-excitation effects are much more complicated than just a reduction of the cross section 
according to the respective SF.
Thus, in this regime one really needs to perform full calculations with the core
excitation and should not rely on a single-particle cross section to
extract the SF. For example, we found that at 31.5 MeV/nucleon beam energy 
the SF extracted in this naive way would be about 70\% too small for the $\frac12^+$ state
but 80\% too large  for the $\frac52^+$ state. This also demonstrates that core
excitation acts very differently in the $S$ and $D$-wave neutron states. 
In the $S$-wave case the results are qualitatively consistent with previous findings
for reactions involving the $\A{11}{Be}(\frac12^+)$ but based on the rotational model.

Taking into account also the study of the $\A{21}{O}^*(p,d)\A{20}{O}(2^+)$ reaction, we are able to make
an important conclusion on a systematic effect of the quadrupole core excitation 
at higher energies: it substantially suppresses reactions with $\ell=0$ transfer 
but enhances those with $\ell=2$. The shape of the angular distribution of the differential cross section
is changed in both cases.
 Of course, the quantitative size of these effects depends on the
collision, binding, and excitation energies.
Furthermore, the core-excitation effect is a result of a complicated interplay between its 
contributions of the two- and three-body nature; 
including only the two-body effect through the modification of the potential is computationally
simpler but not justified.


This work was supported by Lietuvos Mokslo Taryba
(Research Council of Lithuania) under Contract No.~MIP-094/2015.
A.D. acknowledges also the hospitality of the Ruhr-Universit\"at Bochum
where a part of this work was performed.

\vspace{1mm}



\end{document}